\documentclass[journal]{IEEEtran}

\usepackage{epsfig}

\newtheorem{algorithm}{\textit{Algorithm}}
\begin{document}
\title{A Class of Novel STAP Algorithms Using Sparse Recovery Technique}
\author{Hao~Zhang, Gang~Li, Huadong~Meng
\thanks{Submitted April 20, 2009; }
\thanks{Hao Zhang, Gang Li and Huadong Meng are with the Department of Electronics Engineering, Tsinghua University.}}
%

\maketitle

\begin{abstract}
A class of novel STAP algorithms based on sparse recovery technique
were presented. Intrinsic sparsity of distribution of clutter and
target energy on spatial-frequency plane was exploited from the
viewpoint of compressed sensing. The original sample data and
distribution of target and clutter energy was connected by a
ill-posed linear algebraic equation and popular $L_1$ optimization
method could be utilized to search for its solution with sparse
characteristic. Several new filtering algorithm acting on this
solution were designed to clean clutter component on
spatial-frequency plane effectively for detecting invisible targets
buried in clutter. The method above is called CS-STAP in general.
CS-STAP showed their advantage compared with conventional STAP
technique, such as SMI, in two ways: Firstly, the resolution of
CS-STAP on estimation for distribution of clutter and target energy
is ultra-high such that clutter energy might be annihilated almost
completely by carefully tuned filter. Output SCR of CS-STAP
algorithms is far superior to the requirement of detection;
Secondly, a much smaller size of training sample support compared
with SMI method is requested for CS-STAP method. Even with only one
snapshot (from target range cell) could CS-STAP method be able to
reveal the existence of target clearly. CS-STAP method display its
great potential to be used in heterogeneous situation. Experimental
result on dataset from mountaintop program has provided the evidence
for our assertion on CS-STAP.
\end{abstract}


\section{Introduction}
\PARstart{S}pace-time adaptive processing (STAP) is a signal
processing technique that was originally developed for detecting
slowly moving targets using airborne radars\cite{zhang1}. In this
decade, the need for STAP to perform well in heterogeneous
environments, that is, lack of (wide-sense) stationarity of the
received signals with respect to range and snapshots is becoming
more and more urgent\cite{zhang2}. In fact, the unknown clutter and
interference covariance matrix should be estimated from a set of
independent identically distributed (i.i.d.) target-free training
data, which is assumed to be representative of the interference
statistics in a cell under test. The consistency of estimator
depends heavily on the stationarity of training data. In case the
stationary hypothesis is violated, the performance of STAP filter
based on sample covariance matrix of data from training cells will
degrade dramatically. But in practical scenario, stationarity of
training environment tends to disappear when terrain deviates from
being flat with uniform reflectivity properties and in the presence
of internal clutter motion such as tree leaves moving in the wind;
On the other hand, the training data are subject to contamination by
discrete scatterers or interfering targets. radar system
configurations such as conformal arrays and bi-static geometries is
also important factor of non-homogeneity of training data. So the
stationarity is hard to be guaranteed and conventional covariance
estimation methods usually could not be helpful because it is
impossible to get enough ideal training data to calculate the
optimal space-time filtering weight vector.

To overcome the difficulty brought by non-homogeneous clutter and
interference, various methods is proposed to improve the quality of
estimation of covariance matrix. Algorithms of non-homogeneity
detection pay attention to how to detect and eliminate outlier in a
variety of dense target
environments\cite{zhang3}\cite{zhang4}\cite{zhang5}\cite{zhang6}.Methods
in this context is to excise outliers from the training data and use
the resulting outlier-free training data for covariance matrix
estimation But most methods of outlier removal relies on
full-dimension STAP processing and therefore is not suited for
conditions of limited sample support. Especially when main source of
non-stationarity is the great diversity of clutter reflecting
characteristic in training cells, simply detecting and removing the
outlier is not enough to build satisfactory STAP filter.
$D^3$(Direct Data Domain) methods is different from conventional
statistical techniques based upon the covariance
matrix\cite{zhang7}\cite{zhang8}\cite{zhang9}.It use data from the
primary range cell only, and so bypass the problem of the required
homogeneous secondary data support. Although $D^3$ algorithms
perform well when blinking jammer are encountered, they are not as
effective against the homogeneous component of the interference
because they ignore all statistical information. In recent years the
approach received much attention is so called knowledge based
STAP\cite{zhang10}. Traditional STAP methods make little use of a
priori knowledge such as anticipated structure of surface clutter
returns and cultural information available from land-use databases
and digital terrain elevation data (DTED) models\cite{zhang11}. On
the contrary, these prior knowledge is utilized skillfully by
knowledge-based STAP(KA-STAP) techniques to obtain the better effect
of clutter migration. Ordinary KA-STAP methods could be roughly
divided into two category\cite{zhang10}: Intelligent Training/Filter
Selection and Bayesian Filtering/Data Pre-Whitening. Results of
theoretical analysis and experimental evaluation on real data
collected in research program such as MCARM or KASSPER shows that
including knowledge about the operating environment in the signal
processing solution offers significant detection performance
gains\cite{zhang12}\cite{zhang13}. It should be noticed that the
exact form of prior knowledge is problem-dependent and hard to be
derived. Furthermore, how to make effectively use of prior knowledge
in STAP framework is still a problem worth seriously considering.

Over the last decade, the signal processing community spurred by the
work of Donoho\cite{zhang14},Candes and
Tao\cite{zhang15}\cite{zhang16} pay much attention to the field of
Compressed Sensing. Although the its name , especially its
abbreviation "CS", is somewhat misleading, the essence of this
active field is very natural and intuitive: If our target has some
characteristic of "Sparsity" and the process of measurement is
linear, only a few of sample data is required to recover the
original target exactly or approximatively. That is, if the
existence of "Sparsity" could be definite, a great deal of difficult
inverse problem in electrical engineering discipline which have no
effective solver in framework of traditional signal processing could
be solved accurately with some powerful tools such as $L_1$
regularization and corresponding numerical schemes\cite{zhang17}.
Application of Compressed Sensing on Radar is becoming more and more
notable recently and much work focus on the subfield of radar
imaging\cite{zhang18}\cite{zhang19}\cite{zhang20}. The basis for
applicability of Compressed Sensing technique in radar imaging is
there exist apparent feature of "Sparsity" in radar reflectivity of
some natural and artificial scenario, which is measuring target of
radar imaging. It should be mentioned that "Sparsity" can be
revealed not only in radar echo data of time domain, but also from
other viewpoint. For instance, the energy distribution of echo of
clutter on spatial-frequency plane is obvious "Sparse" because most
energy of clutter concentrates on a small region
typically\cite{zhang1}. It provides a great possibility to
estimating the energy in each spatial-frequency cell directly from
sample data with sparse recovery method. That is the preliminary
idea of our work.

In this paper we propose a novel approach to detect the moving
target in strong clutter situation using airborne radar with array
antenna based on sparse processing technique. We called it CS-STAP.
CS-STAP has two advantages. Firstly, data from much fewer range
cells compared with conventional STAP methods based on covariance
matrix estimation is used. So it can avoid in great extent the
influence of heterogeneous clutter on performance of detector. In
some special case, even data from one range cell is sufficient for
us to detect target effectively. Secondly, our approach has ability
of super-resolution. Hence the accuracy of estimation for Doppler
frequency and azimuth angle of target is much higher than other
covariance based STAP schemes. Ability of super-resolution presented
CS-STAP capability of clearing the clutter almost completely and
obtaining massive improvement on SCR. Experimental results on real
data will verify the assertion above.

The paper is organized as follows. In Section II we state the basic
signal model of STAP radar echo data from viewpoint of sparse
recovery. In Section III we give a brief review of necessary
background on Compressed Sensing and some key results will be
described. In Section IV our main algorithms of detection are
proposed. In Section V the performance of new methods is showed with
the real data from Mountain-Top program. Section VI provides a
summary of our contributions.

\section{Signal Model from viewpoint of Sparse
Processing}\label{sec1}

Considered here is a radar system that transmits coherent pulse
trains with length $L$ and samples the returns on an $N$-element
uniform linear array(ULA). For each pulse, It collects $M$ temporal
samples from each element receiver, where each time sample
corresponds to a range cell. The temporal dimension of interest here
is referred to as fast time. The collection of samples for $m$th
range cell is represented by an $N{\times}L$ data matrix $X_m$ with
elements $X_m(n,l)$, and the pulse dimension of interest here is
referred to as slow time. echo data from all range cells is arranged
into a "Data Cube", that is:
\begin{equation}
X=[X_1,X_2,\cdots,X_L]^T,
\end{equation}
which is the data basis for almost all STAP processing.

Suppose the amplitude of radar echo signal in the same coherent
processing interval(CPI) from a fixed range cell, losing no
generality, $k$th range cell, is stationary, that is, it doesn't
have any fluctuation, then the spatial information (azimuth angle of
echo) and Doppler information (moving speed with respecting to
ground) of object in observing region of radar are all contained in
phase difference among elements of data matrix corresponding to that
range cell.

Let $d$ equal the element spacing of ULA and place the first antenna
element at the origin and designate it as phase reference. The
spatial phase shift on received data at different elements is
\begin{equation}
\phi_s=[1,\exp(j2{\pi}f_s),\cdots, \exp(j2{\pi}\cdot(N-1)f_s)],
\end{equation}
which is usually referred as spatial steering vector, where the
spatial frequency is
\begin{equation}
f_s=\frac{d}{\lambda}\cos(\theta)
\end{equation}
$\lambda$ is wavelength of radar and $\cos(\theta)$ is directional
cosine.

On the other hand, suppose motion of scatterer, motion of radar
platform, or both lead to Doppler frequency $f_d$
\begin{equation}
f_d=\frac{2v}{\lambda},
\end{equation}
where $v$ denotes the radical velocity component, or line-of-sight
velocity. Let the first pulse in one CPI be the reference point, we
obtained Doppler steering vector, the representation of Doppler
phase shift on received data of different pulses in this CPI.
\begin{equation}
\phi_d=[1, \exp(j2{\pi}f_d),\cdots, \exp(j2{\pi}\cdot(L-1)f_d)],
\end{equation}

The Spatial-Doppler steering vector corresponding to the echo signal
from a point scatterer with spatial frequency $f_s$ and normalized
Doppler frequency $f_d$ is
\begin{eqnarray}
\phi_{s-t}(f_s,f_d)&=&\phi_s\otimes\phi_d\nonumber\\
&=&[1,\exp(j2{\pi}f_d),\cdots,\exp(j2{\pi}(L-1)f_d),\nonumber\\
&&\cdots,\exp(j2{\pi}((N-1)f_s+(L-1)f_d))],\nonumber\\
\end{eqnarray}
where $\otimes$ denotes Kronecker product. Assume the contribution
of this point scatterer in overall radar echo is $\alpha(f_s,f_d)$
which is a complex quantity, then the echo signal snapshot from
fixed range cell can be obtain by summarizing the contribution from
individual point scatterer with different spatial and Doppler
frequency together,
\begin{equation}
r=\sum_{f_s,f_d}\alpha(f_s,f_d)\phi_{s-t}(f_s,f_d),
\end{equation}
Here $r$ is a vector with the same length as $\phi_{s-t}(f_s,f_d)$,
that is, $N{\times}L$.

Discretizing spatial and Doppler axis, we divided the
spatial-frequency plane into square grids. The content in each grid
represents the echo signal with spatial frequency and Doppler
frequency corresponding to that grid. Then the data snapshot could
be written as
\begin{equation}
r=\sum_{m=1}^{N_s}\sum_{n=1}^{N_d}\alpha(m,n)\phi_{s-t}(m,n),\label{label1}
\end{equation}
where $N_s$ and $N_d$ is the number of quantization of spatial and
Doppler axis respectively. Roughly speaking, they can be set
arbitrarily. But we must make carefully choice on $N_s$ and $N_d$ to
build practical algorithm.

It should be noted that model (\ref{label1}) is different from the
data model proposed in \cite{zhang21} as follows:
\begin{equation}
r=\sum_{m=1}^{N_a}\sum_{n=1}^{N_c}\beta(m,n:k)\xi_{s-t}(m,n:k),
\end{equation}
where $N_a$ and $N_c$ denotes the number of statistically
independent clutter patches in iso-range and ambiguous ranges of
$k$th range respectively. In our model, the range ambiguity is not
considered and the independence of clutter in distinct grid is not
required (Although it is meaningful in our algorithm design). model
(\ref{label1}) can be written as matrix form
\begin{equation}
r=\Phi{x},\label{label2}
\end{equation}
where
\begin{eqnarray}
\Phi&=&[\phi_{s-t}(1,1),\cdots,\phi_{s-t}(1,n),\cdots,\phi_{s-t}(m,n)],\label{label6}\\
x&=&[\alpha(1,1),\cdots,\alpha(1,n),\cdots,\alpha(m,n)],
\end{eqnarray}

Equation (\ref{label2}) is the fundamental model in this paper. It
has three important characteristic worthy of indicating explicitly.
at first, the energy distribution of clutter and target buried in it
on spatial-frequency plane could be obtained by solving linear
equation (\ref{label2}) of unknown $x$ and known $r$ which is
spatial-temporal sample data. In other words, operation of STAP is
essentially a process of solving linear algebraic equation from our
point of view. Secondly, the row dimension $N{\times}L$ of matrix
$\Phi$ is much less than column dimension $N_s{\times}N_d$
typically. Hence equation (\ref{label2}) is heavily ill-posed. The
number of unknown variables is much more than the number of
equations. That is to say, there are infinite vectors satisfying
(\ref{label2}) in general and some constraint should be imposed to
help us to get some solution with unique feature. Thirdly, from
viewpoint of STAP, the solution of (\ref{label2}) we are searching
for is "Sparse". More rigorously, most of its entry are negligible
and only a small portion of elements in solution vector, which
represents contribution of main clutter and target, are remarkable,
just as description in standard textbook of STAP
\cite{zhang1}\cite{zhang2}. This kind of "Sparsity" is the basis of
application of sparse recovery technique on STAP. Before our new
approach is proposed, some necessary background of sparse processing
will be sketched briefly.

\section{a brief review of necessary background on Compressed Sensing}

One of the main objective of compressed sensing is seeking the
solution with "Sparse" property of heavily ill-posed linear
algebraic equations (\ref{label3})
\begin{equation}
y=Ax,\label{label3}
\end{equation}
where $A\in{C}^{M{\times}N}$ is a "rectangle" matrix, that is,
$M{\ll}N$. Of course, equation (\ref{label3}) has infinite solutions
if no further assumption is put forward. In many practical situation
(including STAP), people focus their interests on the most "Sparse"
solution of (\ref{label3}), that is, the one with minimal number of
"nonzero" elements. Intuitively, the process of solving equation
(\ref{label3}) for the most "Sparse" solution can be transformed to
so called $L_0$ optimization problem as follows:
\begin{equation}
\min\|x\|_0, {\quad}s.t{\quad}Ax=y,\label{label4}
\end{equation}
or more flexibly
\begin{equation}
\min\|x\|_0, {\quad}s.t{\quad}\|Ax-y\|_2\leq\epsilon,
\end{equation}
This is a combinatorial optimization problem and can be proven to be
NP-hard\cite{zhang22}. Hence $L_1$ relaxation is utilized as
substitution of intractable $L_0$ problem to find the most "Sparse"
solution. Specially, our optimization problem is changed to
\begin{equation}
\min\|x\|_1, {\quad}s.t{\quad}Ax=y,\label{label5}
\end{equation}
or more flexibly
\begin{equation}
\min\|x\|_1, {\quad}s.t{\quad}\|Ax-y\|_2\leq\epsilon,
\end{equation}
where $\|x\|_1=|x_1|+\cdots+|x_n|$ for $x\in{C}^n$. (\ref{label5})
is a linear programming problem essentially and can be solved
effectively by popular convex optimization algorithm \cite{zhang23}
or greedy method \cite{zhang24}.

A important question was raised naturally: whether (\ref{label4})
and (\ref{label5}) is really equivalent? Candes and Tao defined a
kind of character property for matrix $A$ named Restricted Isometry
Property (RIP)\cite{zhang16} in order to study $L_0/L_1$
equivalence. RIP concerns with the upper and lower bound of singular
values of the submatrix of $A$. In particular, for every $K\in{N}$,
matrix $A$ is said to satisfy RIP of order $K$ if there exist a
positive constant $\delta_K$, such that
\begin{equation}
(1-\delta_K)\|z\|_2^2\leq\|Bz\|_2^2\leq(1+\delta_K)\|z\|_2^2,
\end{equation}
for all submatrix of $A$ with dimension $M{\times}K$, where
$z\in{C}^K$. in other words, the singular values of all submatrix of
$A$ with dimension $M{\times}K$ should be restricted in interval
$(1-\delta_K,1+\delta_K)$. Several sufficient conditions based on
the value scope of RIP constant of different order have been
proposed to guarantee the $L_0/L_1$ equivalence, For
example\cite{zhang16},
\begin{equation}
\delta_{3K}+3\delta_{4K}<2,
\end{equation}
and\cite{zhang25}
\begin{equation}
\delta_{2K}<\sqrt{2}-1,
\end{equation}

Although many methods to construct matrix satisfying RIP(especially
randomization method) have been brought forward\cite{zhang17}, a
lack of effective means for testing RIP and estimating RIP constant
of a given deterministic matrix is still a hidden trouble for
practitioners and user of compressed sensing
technique\cite{zhang26}, no exception for us. It is very hard to
verify RIP of matrix used in STAP operation based on sparse recovery
algorithm theoretically. So the detailed conditions for $L_0/L_1$
equivalence will not be checked in our research. Fortunately, no
severe consequence was encountered generally, because simply using
$L_1$ optimization could we find the solution of (\ref{label3}) with
some characteristic of sparsity (although it maybe isn't the most
sparse one). That is just what we want in most cases.

\section{STAP Algorithms Based On Sparse Recovery}\label{sec2}

It is easy to notice that linear equation (\ref{label1}) and
(\ref{label3}) have some similarity: the row dimension of
coefficient matrix is much smaller than its column dimension; the
target solution is "Sparse". So it's reasonable to transform
equation (\ref{label1}) into a $L_1$ optimization problem to obtain
the solution via convex or greedy method.

In practice, the space-time covariance matrix which is key device in
framework of traditional STAP is not known in prior and needs to be
estimated from the secondary data. This is just the point for which
Sparse recovery method can take advantage of to improve their
performance when sample support of snapshots is tiny. The reason is
 the contribution of point scatterer with specific spatial and
Doppler frequency in overall echo could be estimated
\textbf{directly} from snapshot by sparse recovery algorithm. So the
spatial-frequency distribution of clutter could be calculated with
few snapshots. Then it could be used to design effective filter to
eliminate clutter and make the detection for small target possible.

Using the notation and abbreviation in Section \ref{sec1}, let
$r\in{C}^{N{\times}L}$ be sample data from single snapshot of
preliminary data. In other words, the data we will deal with come
from cell under test only. it may contains both clutter and target.
No training cells are used in our treatment of single snapshot. From
this point, the algorithm below belongs to $D^3$ category.

\mbox{}

\begin{algorithm}[Annihilating Filter --- Single Snapshot
Case]\label{label17}

\mbox{}

\begin{itemize}
\item[1] Choose the row dimension $N_s$ and column dimension $N_d$ for
quantization grids of spatial-frequency plane.

\item[2] Form measurement matrix $\Phi$ with $N{\times}L$ rows and
$N_s{\times}N_d$ columns as (\ref{label6}).

\item[3] Solve linear ill-posed equation $r=\Phi{x}$ to obtain $x$,
the estimation of levels of echo signal with given
spatial and Doppler frequency.

\item[4] Calculate the absolute value of each entry in $x$ to get
$|x|=(|x_1|,\cdots,|x_{N_s{\times}N_d}|)$.

\item[5] Arrange the entries of $|x|$ in descend order, and estimate
the index $k$ such that
\[
|x_{[1]}|>|x_{[2]}|>\cdots>|x_{[k]}|\gg|x_{[k+1]}|>\cdots>|x_{[N_s{\times}N_d]}|,
\]

\item[6] Set $|x_{[1]}|=|x_{[2]}|=\cdots=|x_{[k]}|=0$ and output
$|x|$.
\end{itemize}

\end{algorithm}
\mbox{}

If Clutter-Noise-Ratio(CNR) and Clutter-Signal-Ratio(CSR) are both
sufficient large, then algorithm \ref{label17} can eliminate the
clutter energy in sample data efficiently. we can use conventional
threshold detection on the output of algorithm \ref{label17} to
reveal the existence of target. Algorithm \ref{label17} is a kind of
"whiten" filter.

Except for solving linear ill-posed equation, the key step in
algorithm \ref{label17} is step (5), that is, find the accurate
position of clutter on spatial-frequency plane. The assumption on
CNR and CSR is critical in this step. In fact, if CNR wasn't large
enough, then it was hard to find the correct index $[k]$ that can
separate clutter from noise. This oracle has an analogy to that of
standard MUSIC algorithm for super-resolution spectral estimation,
where the right separation of signal subspace and noise subspace
relies on finding the index where the eigenvalues of sample
covariance matrix with descend order have a abrupt change. if SNR
wasn't large adequately, eigenvalues would change gradually and the
performance of estimator of noise subspace degraded dramatically.
Hence MUSIC algorithm will become invalid. Here the problem is
alike. Secondly, if CSR, instead of CNR, is too small, the target is
easy to be taken in mistake as part of clutter and be removed from
data by Annihilating operation. On the other hand, because only one
data snapshot is used and there is no any integration in algorithm
\ref{label17}, the noise receives no suppress and the algorithm is
statistically unstable.

To increase CNR and improve the statistical stability of algorithm
\ref{label17}, integration of multiple data snapshots with respect
to different range cells is added. It should be emphasized that
although data from training cells are used in this algorithm, the
necessary sample support is much smaller than that of standard STAP
filter based on covariance matrix in practice.

Suppose sample support for our algorithm is $K$, Let $r^{(i)}$ be
$i$th snapshot from training cell, $i=1,2,\cdots,K$, $r^{[T]}$ be
snapshot from testing cell, then we have

\mbox{}

\begin{algorithm}[Annihilating Filter --- Multiple Snapshots
Case]\label{label7}
\mbox{}
\begin{itemize}
\item[1] Choose the row dimension $N_s$ and column dimension $N_d$ for
quantization grids of spatial-frequency plane.

\item[2] Form measurement matrix $\Phi$ with $N{\times}L$ rows and
$N_s{\times}N_d$ columns as (\ref{label6}).

\item[3] Solve linear ill-posed equations to obtain $x^{(k)}$ and
$x^{[T]}$ corresponding to $\{r^{(i)}\}$ and $r^{[T]}$ respectively.

\item[4] Estimate the statistical mean of $(x^{(1)},\cdots,x^{(K)})$
by
\[
x=\frac{x^{(1)}+x^{(2)}+\cdots+x^{(K)}}{K},
\]

\item[5] Calculate the absolute value of each entry in $x$ to get
$|x|=(|x_1|,\cdots,|x_{N_s{\times}N_d}|)$.

\item[6] Arrange the entries of $|x|$ in descend order, and estimate
the index $k$ such that
\[
|x_{[1]}|>|x_{[2]}|>\cdots>|x_{[k]}|\gg|x_{[k+1]}|>\cdots>|x_{[N_s{\times}N_d]}|,
\]
record the position $([1],[2],\cdots,[k])$.

\item[7] Set $|x^{[T]}_{[1]}|=|x^{[T]}_{[2]}|=\cdots=|x^{[T]}_{[k]}|=0$ and
output $|x^{[T]}|$
\end{itemize}

\end{algorithm}
\mbox{}

Through calculating the sample mean, the influence of zero-mean
thermal noise will be controlled evidently. Further, some simple
robust technique, such as censoring or compute sample median instead
of mean, could bring some advantage on impulse noise including
flicker or artificial interference.

There are more improvement for algorithm \label{label7}.
Experimental result reminds that simply "Annihilate" the big
(Clutter) component on spatial-frequency plane is not enough for
revealing target, because equation (\ref{label2}) is heavily
ill-posed. More specifically, there exists some strong correlation
between the columns of matrix $\Phi$ (In fact, systematically study
of solvable property, such as RIP, of matrix $\Phi$ consisting of
spatial-temporal steering vectors used in STAP hasn't been conducted
until now). So apparent error may appears in the solution of
(\ref{label2}). The influence of clutter scatterers with specific
spatial and Doppler frequency spread to not only adjacent cells, but
also to somewhere far away on spatial-frequency plane. Intuitively
speaking, this is some kind of "sidelobe" or "Pseudo-Peak". Although
we argued that STAP based on compressed sensing has
"super-resolution" property, the sidelobe is unavoidable for the
intrinsic constraint imposed by matrix $\Phi$.

Migration of "sidelobe" in great extent could be achievable as
follows: Determine the positions of big components on
spatial-frequency plane thought the training data.  Set the
corresponding entries of solution vector of (\ref{label2}) on data
from testing cells to zero. Furthermore, the entries of solution
vector with respecting to columns of matrix $\Phi$ which have large
correlation with that of clutter scatterers is also set to zeros.

\mbox{}

\begin{algorithm}[Sidelobe Suppressing Filter]\label{label8}
\mbox{}
\begin{itemize}
\item[1] Choose the row dimension $N_s$ and column dimension $N_d$ for
quantization grids of spatial-frequency plane.

\item[2] Form measurement matrix $\Phi$ with $N{\times}L$ rows and
$N_s{\times}N_d$ columns as (\ref{label6}).

\item[3] Solve linear ill-posed equations to obtain $x^{(k)}$ and
$x^{T}$ corresponding to $\{r^{(i)}\}$ and $r^{[T]}$ respectively.

\item[4] Estimate the statistical mean of $(x^{(1)},\cdots,x^{(K)})$
by
\[
x=\frac{x^{(1)}+x^{(2)}+\cdots+x^{(K)}}{K},
\]

\item[5] Calculate the absolute value of each entry in $x$ to get
$|x|=(|x_1|,\cdots,|x_{N_s{\times}N_d}|)$.

\item[6] Find the maximal entry $|x_{M}|$ in $|x|$ with index
$M$, and corresponding column $\phi_{M}$ in matrix $\Phi$.

\item[7] Set $|x^{[T]}_{k}|=0, k\in{K}$ and $|x_{k}|=0, k\in{K}$ where
\[
K=\{k\in{N}: |<\phi_k, \phi_{M}>|\geq\Delta\}
\]
where $\Delta$ is a prescribed threshold value.

\item[8] calculate the residue energy of $|x|$,
if it is lower than constant $\epsilon$, output $|x^{[T]}|$; else go
to step (6).
\end{itemize}

\end{algorithm}
\mbox{}

It should be remarked that two constant, $\Delta$ and $\epsilon$, is
critical to performance of algorithm \ref{label8}. It depend on
clutter scenario and SNR and must be chosen carefully. Bad setting
of constant may result in that target is obscured by sidelobe
leakage from the clutter scatterer or deleted mistakenly in the
process of sidelobe suppression.

\section{Numerical Results}

Numerical results presented in this section were derived from
processing publicly available real data collected by the DARPA
sponsored Mountain Top program\cite{zhang29}. This data was
collected from commanding sites (mountain tops) and radar motion is
emulated using a technique developed at Lincoln
Laboratories\cite{zhang27}. The sensor consists of 14 elements and
the data is organized in CPIs of 16 pulses. For the data set
analyzed here, the clutter was located around 70 degree azimuth at a
Doppler frequency of -150 Hz. A synthetic target was introduced in
the data at 100 degree and -150 Hz. Note that the clutter and target
have the same Doppler frequency, hence separation is possible only
in the spatial domain. The target is not visible without processing.
A $16\times14$ matrix was formed for each range cell, where the
temporal information is in the matrix columns, and the spatial
information is contained in the rows.

Choice of software package for $L_1$ optimization needed further
consideration. There are two category of algorithms for solving
ill-posed linear equations: Greedy pursuit and convex optimization.
The most popular package of convex programming is cvx developed by
S.Boyd and his colleagues\cite{zhang30}. although the robustness is
excellent, efficiency of cvx is not very satisfactory and it isn't
suitable for large scale numerical experiment. Greedy pursuit, such
as OMP\cite{zhang31}, regularized OMP\cite{zhang32} and
CoSaMP\cite{zhang33}, have much faster convergence rate. But
stability is still a problem with most greedy schemes. Especially in
the case that the solution we are searching for is compressible
signal, not strictly sparse signal, effect of recovery using greedy
algorithm becomes unsatisfactory. However, in the field of radar
signal processing, strictly sparse signal doesn't exist at all for
the noise is universal. So there is urgent need for stable greedy
method for application of signal processing and communication. The
one used in this section is designed by ourself\cite{zhang34}. Its
performance was illustrated in numerical experiment.

\subsection{One data snapshot case}

Traditional STAP filter, including SMI and various Reduce Rank
versions, utilized estimation of covariance matrix as the tool for
adaptive processing to obtain the ultra-low taper. Clearly, The
target can't be detected just using unadapted weight vector such as
steering vector, that is, DFT or DCT. However, sparse recovery based
filter can give the estimation of target and clutter directly
without help of covariance matrix. In other words, no training data
is needed in the processing and one data snapshot from testing cell
is sufficient for clutter and target to be visible on
spatial-frequency plane.

\begin{figure}
  \centering
  \centerline{\epsfig{figure=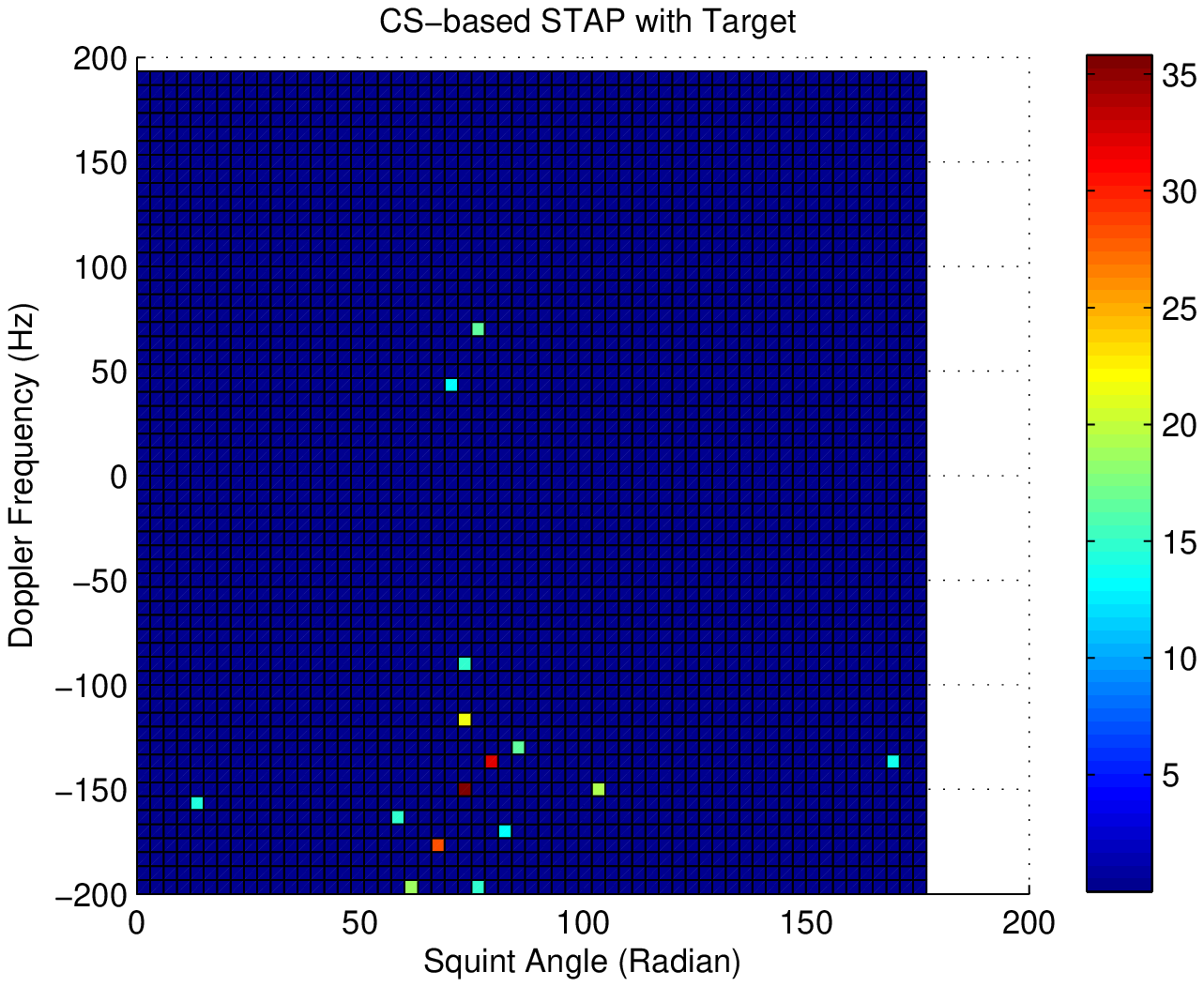,width=10cm}}
  \centerline{A}\medskip
  \centerline{\epsfig{figure=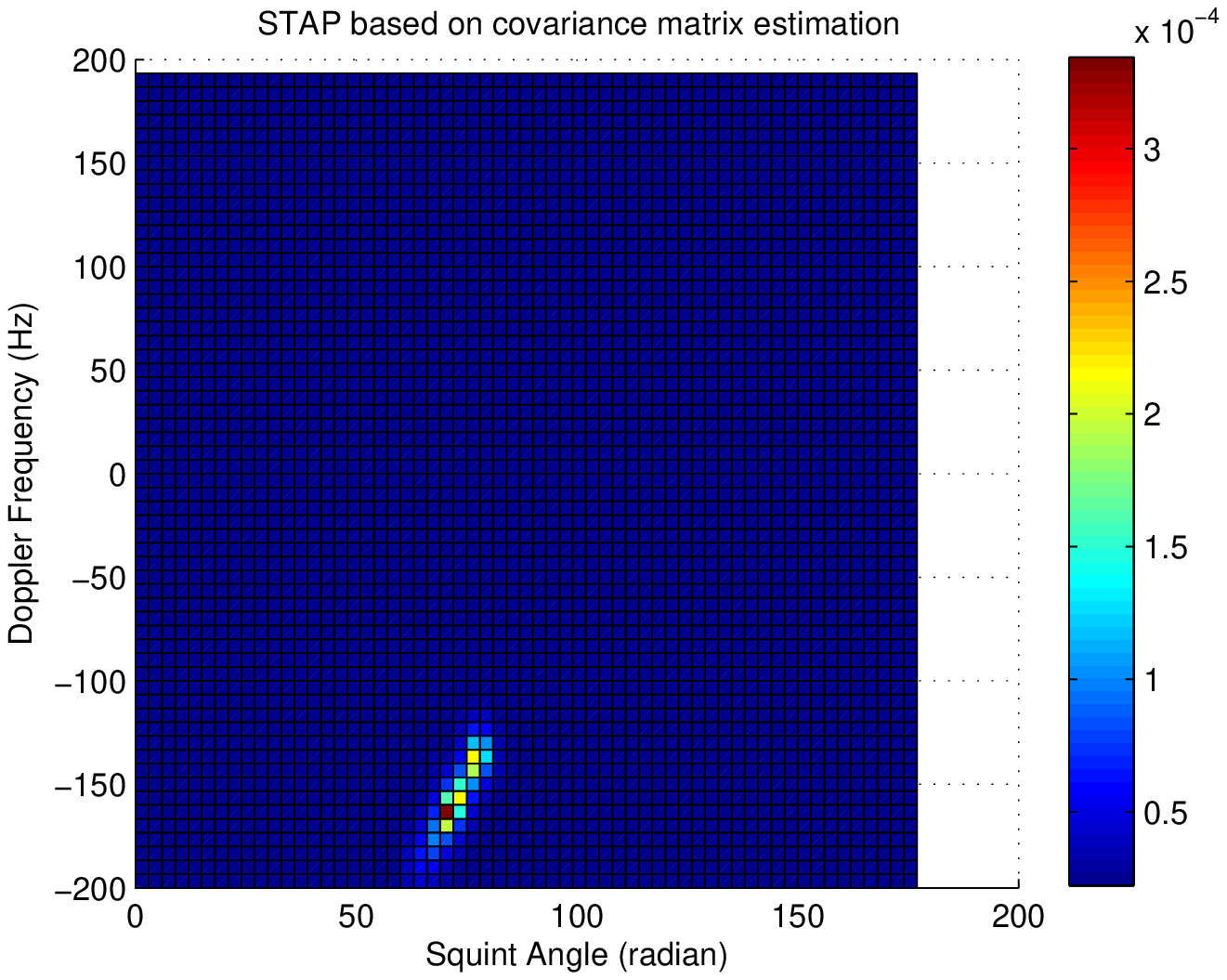,width=10cm}}
  \centerline{B}\medskip
  \caption{Echo amplitude of clutter and target on spatial-frequency
plane} \label{fig1}
\end{figure}

Fig \ref{fig1}(A) showed the estimation of energy distribution on
spatial-frequency plane at the target range cell using algorithm
\ref{label17}. Here x-axis is for azimuth angle and y-axis is for
Doppler frequency. As comparison, corresponding result got by using
SMI(Sample Matrix Inversion) algorithm with only three data
snapshots and diagonal loading was shown in Fig \ref{fig1}(B).

It is clear that estimation based on sparse recovery has much higher
resolution than that of conventional covariance method. Especially
in clutter region, individual scatterers on each resolution cells
with different spatial angles and Doppler frequency could be
distinguished in Fig \ref{fig1}(A), and in Fig \ref{fig1}(B) the
clutter shows itself as a "block". More important, the target which
has normalized Doppler frequency near -150Hz (the clutter center has
almost the same Doppler frequency) and spatial angles near 100
degree, is distinct on Fig \ref{fig1}(A), but it is invisible
absolutely on Fig \ref{fig1}(B). In fact, the target can only be
visible when clutter is eliminated by adaptive processing. It should
be noticed that clutter energy spreads to somewhere far away from
clutter region, just as we mentioned in section \ref{sec2}. This is
the reason of adoption of sidelobe suppression filter and multiple
data snapshots.

\subsection{Multiple data snapshots case}

\begin{figure}
  \centering
  \centerline{\epsfig{figure=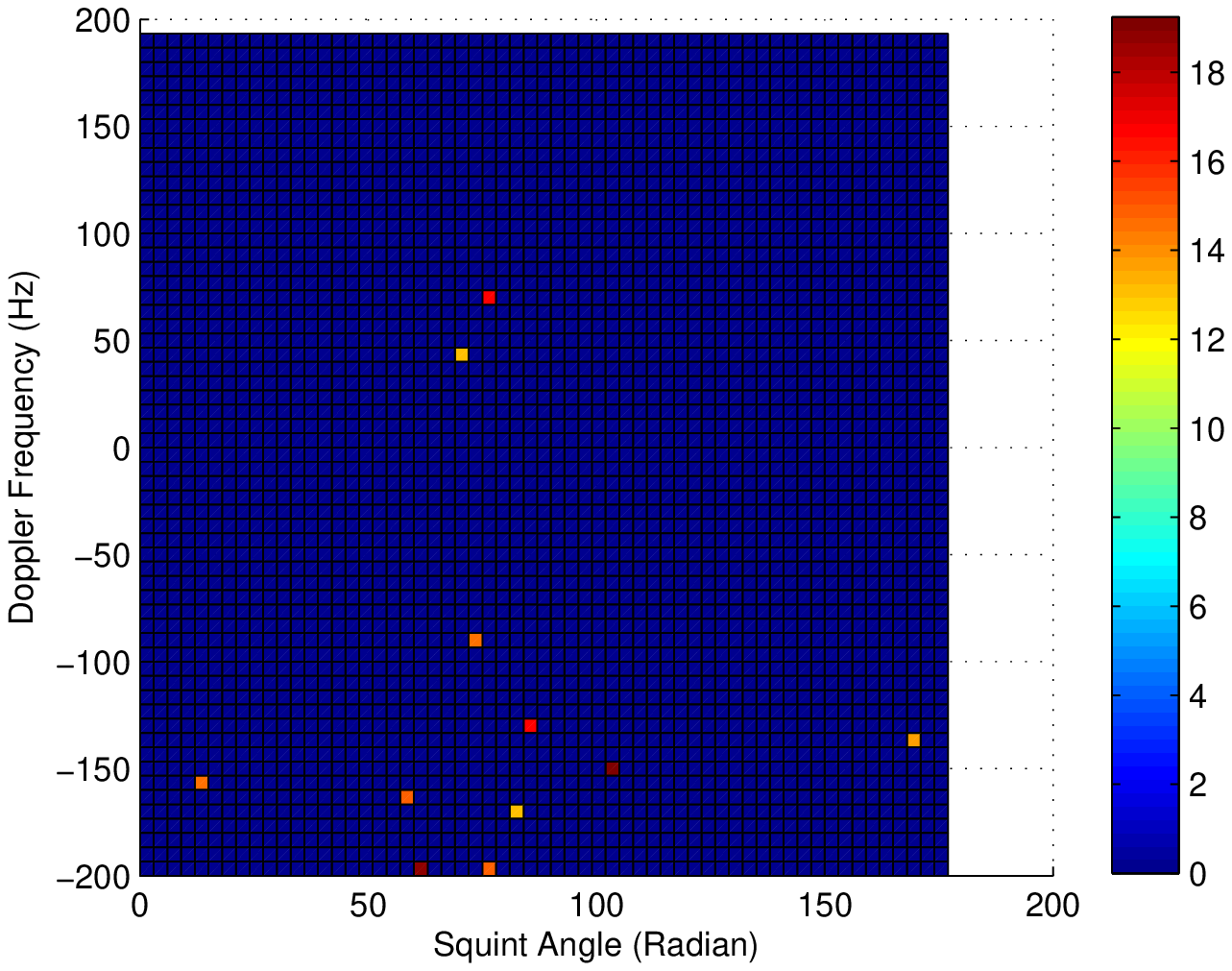,width=10cm}}
  \centerline{A}\medskip
  \centerline{\epsfig{figure=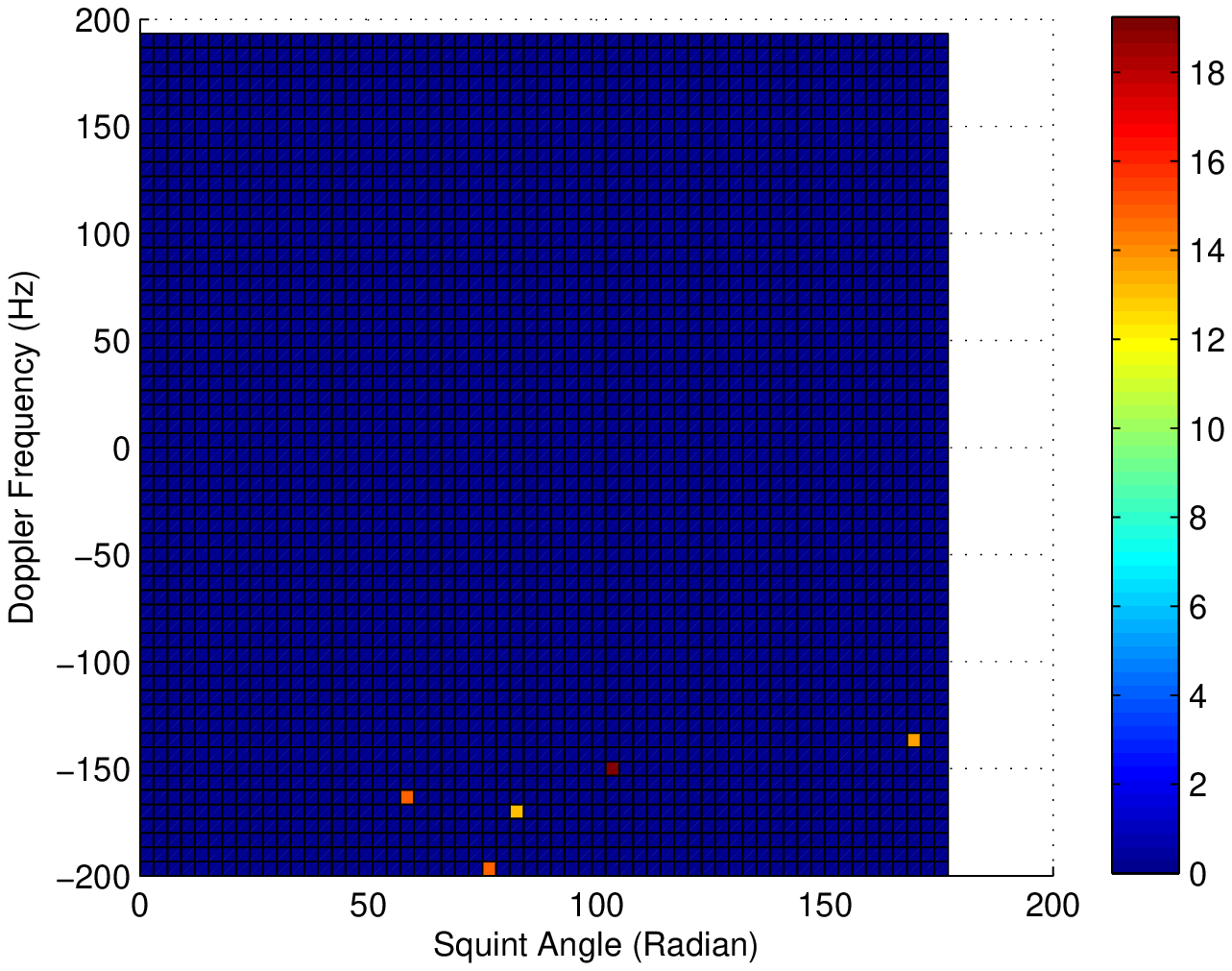,width=10cm}}
  \centerline{B}\medskip
  \centerline{\epsfig{figure=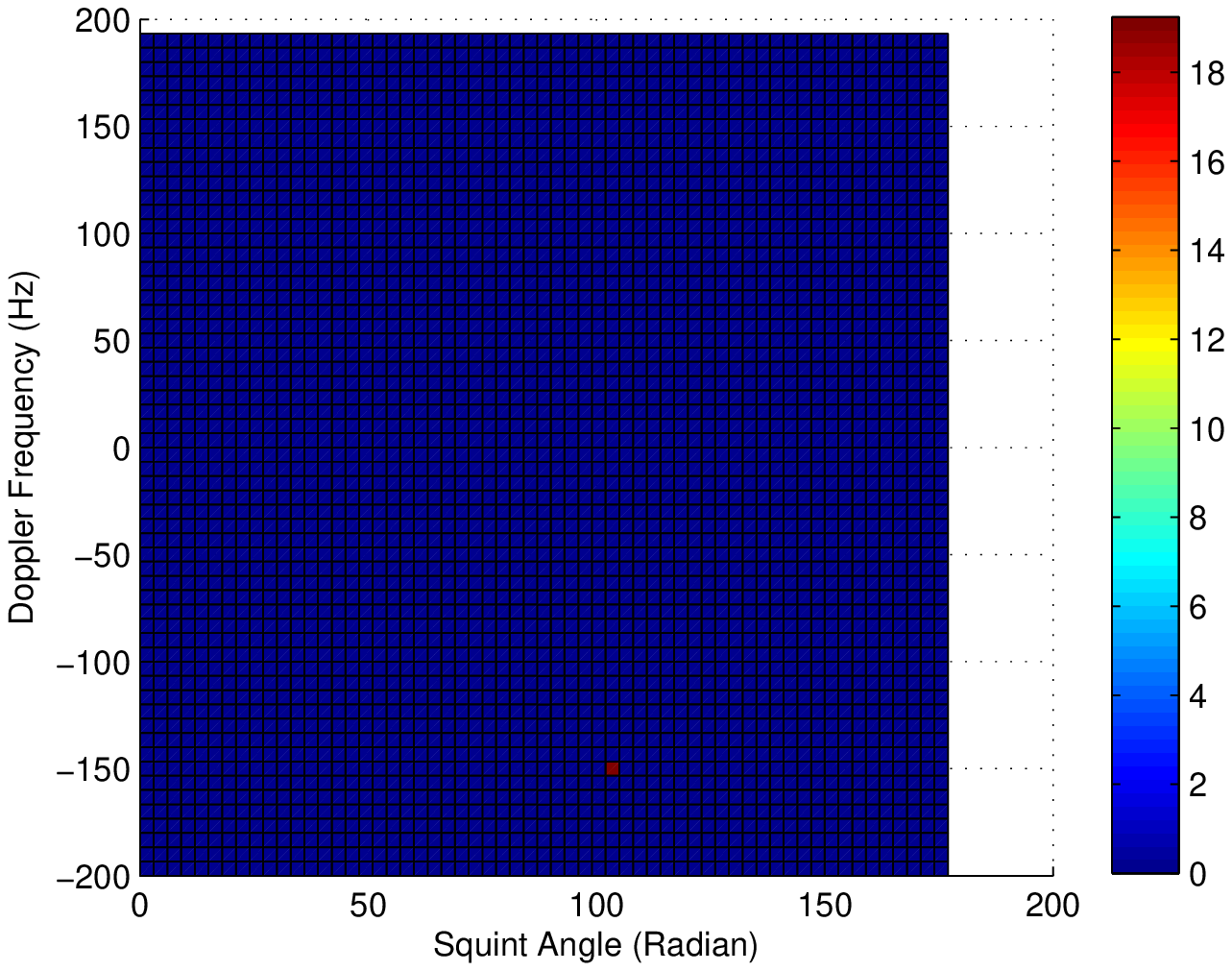,width=10cm}}
  \centerline{C}\medskip
  \caption{Effect of sidelobe suppressing filter} \label{fig2}
\end{figure}

\begin{figure}
  \centering
  \centerline{\epsfig{figure=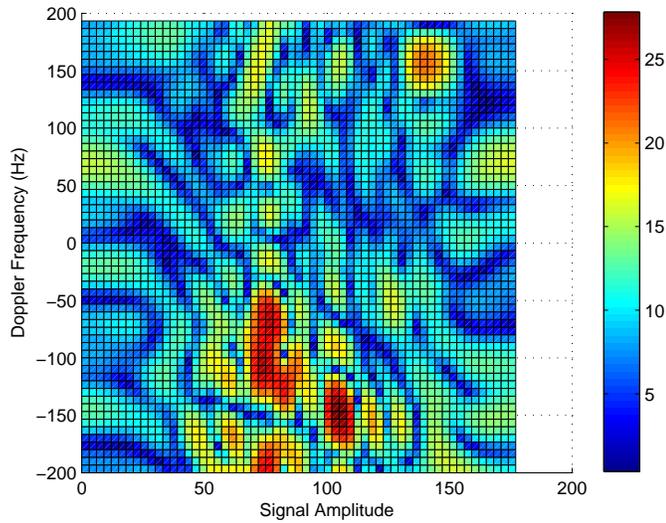,width=10cm}}
  \caption{Effect of SMI filter} \label{fig3}
\end{figure}

Fig \ref{fig2} illustrated the effect of sidelobe suppression
filtering. Training data is necessary in this setting for estimating
the level of clutter. (A), (B) and (C) of figure \ref{fig2} showed
the situation on spatial-frequency plane of target range when 70, 90
and 95 percentages of clutter energy was removed from training data,
as implemented in algorithm \ref{label8}. 16 range cells around
target range with 5 guard range cells are employed for training. The
result of adaptive covariance filtering by SMI algorithm with the
same training data is showed in Fig \ref{fig3}.

The advantage of sidelobe suppressing filter is presented
sufficiently by figure \ref{fig2} and \ref{fig3}. For SMI method ,
shortage of training data led to heavy degradation of performance.
In fact, he size of sample support is much less than DOF (Degree Of
Freedom) of clutter, so the estimator for clutter covariance matrix
is cripple and the leakage of clutter from filter is unacceptable.
Besides that, the resolution of target is so low that it is
impossible to determine the accurate azimuth angle and Doppler
frequency of target. On the other hand, small size of training data
had little influence on the performance of sidelobe suppressing
filter based on sparse recovery because it didn't involve complex
statistical estimation of sample covariance matrix, which had good
asymptotic property only when sample size is sufficient large. The
target and clutter is revealed directly from sample data and
multiple data snapshots is helpful to decrease the effect of noise.
Sidelobe suppressing filter is also preferred for its super-high
resolution. It provides possibility for effectively detecting the
slow targets.

\subsection{Gains at various steering angles}

\begin{figure}
  \centering
  \centerline{\epsfig{figure=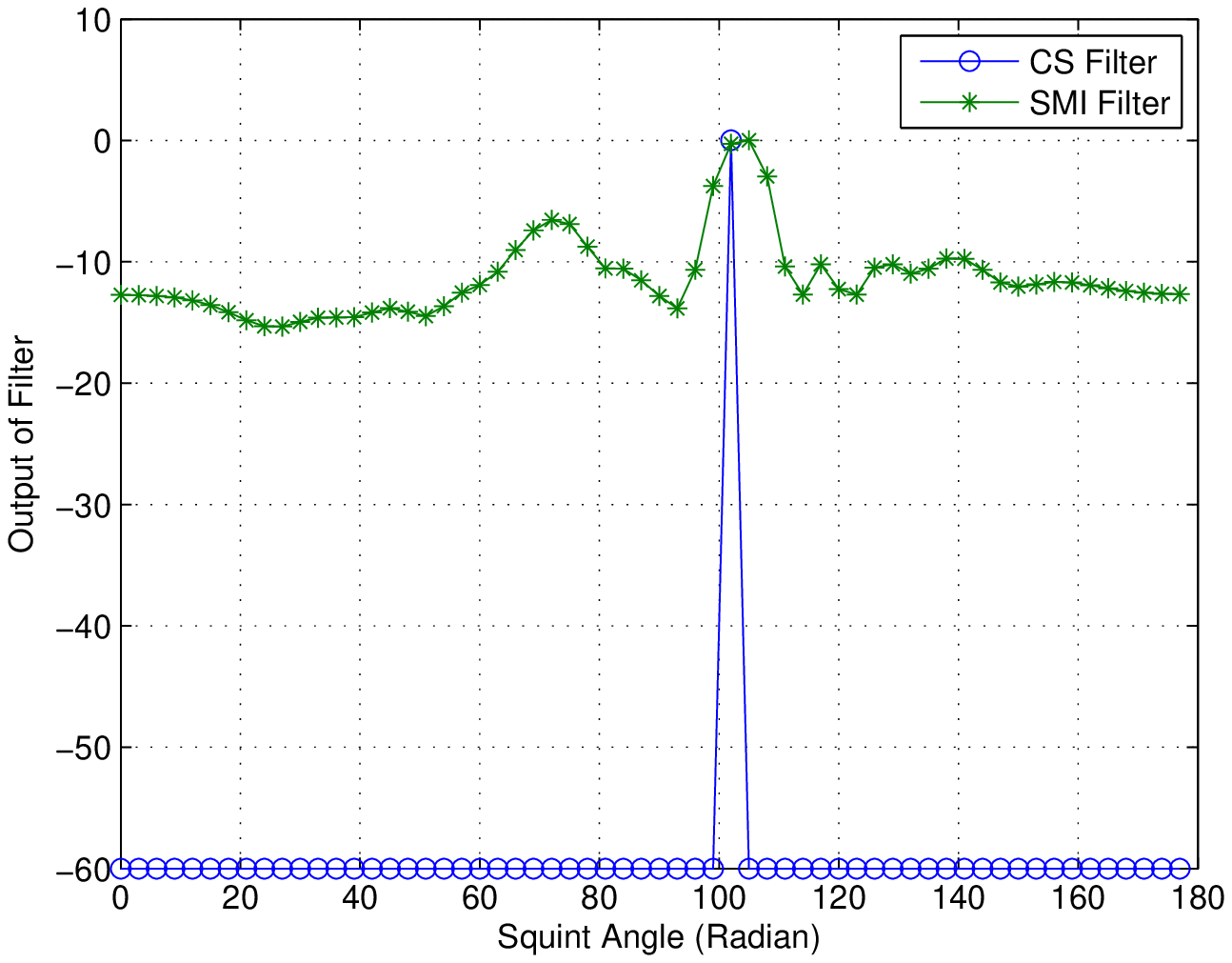,width=10cm}}
  \caption{Output of Filter at Different Squint Angle} \label{fig4}
\end{figure}

Fig \ref{fig4} displays the echo magnitude at the target range cell
at various azimuth angles. The figure was generated by scanning the
steering angle over the angular sector indicated by the abscissa.
The gain of each technique is measured with respect to the output at
the actual target direction (at about 100 degree). This experiment
emulates the search mode of the radar. In the case presented, Near
the target angle, the output of filter based on CS and SMI all reach
a maximum. But output of CS filter is much lower than that of SMI at
other angles except for target, even under the situation that 80
training cells were used in SMI filter but CS filter only used data
from 16 range cells. In fact, CS filter only has output at target
angle because the echo on other angle is regarded as clutter and
eliminated by sidelobe suppressing filter. This characteristic of CS
filter showed its ability of super-resolution on spatial domain. We
argue that technique of CS could be applied in such field as
spectrum estimation and high-resolution direction of arrival (DOA),
for its remarkable potential of super-resolution.

\subsection{Gains at various range cells}

\begin{figure}
  \centering
  \centerline{\epsfig{figure=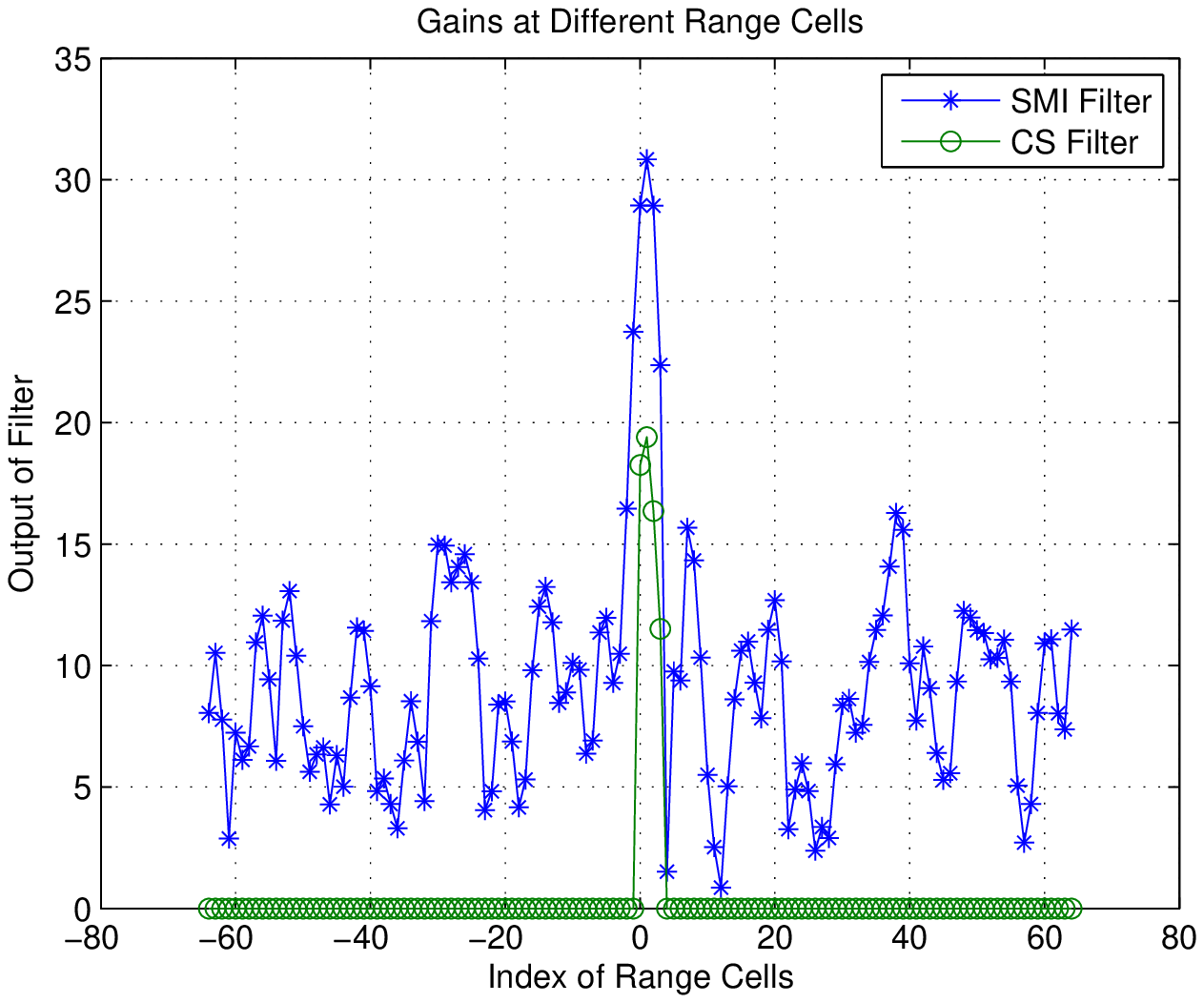,width=10cm}}
  \caption{Output of Filter at Different Squint Angle} \label{fig5}
\end{figure}

Fig \ref{fig5} shows the output of the CS filter and SMI filter at
the range cells for which data is available. The covariance matrix
was estimated from samples from 80 range cells, not including the 5
samples around the target range. Both the SMI and the CS methods
detect the target and suppress the clutter. But CS method output
nothing outside target range because of its ability of
super-resolution for revealing target buried in strong clutter
explicitly and ability of sidelobe suppressing filter to clear out
the clutter when target isn't presented. CS method only used 16
range cells for training, much less than that of SMI method. So it
is more suitable for being applied in non-homogeneous clutter
environment.

\section{Conclusion}

In this paper, a class of novel STAP algorithms based on sparse
recovery technique, called CS-STAP, were given and examined. Its
performance is studied by analysis of real data. The motivation for
application of technique of compressed sensing on STAP is inherent
sparsity of radar echo from target and clutter. Building estimator
directly from sample data make CS-STAP method have ability of
super-resolution and lead further to almost complete elimination of
clutter. Improved performance over conventional STAP method
depending on covariance matrix estimation, such as SMI, were
obtained by CS-STAP method. In an application of this method to the
Mountaintop data set, we showed that CS-STAP method can provide
better effect of clutter suppression and more accurate estimation
for Doppler frequency and azimuth angle of target than SMI detector.
Furthermore, the number of training data used in CS-STAP method is
much less than SMI. It means that CS-STAP method will outperform
significantly SMI method when the hypothesis of stationary for
clutter is violated severely.

%
%

\end{document}